\documentclass[aps,prd,a4paper,preprintnumbers,amsmath,amssymb,superscriptaddress,twocolumn,floatfix,showpacs,showkeys,nofootinbib]{revtex4}
\usepackage{amssymb}

\usepackage{float,graphicx}

\begin{document}

\title{The Cosmology of Modified Gauss-Bonnet Gravity}
\author{Baojiu~Li }
\email[Email address: ]{B.Li@damtp.cam.ac.uk}
\affiliation{Department of Applied Mathematics and Theoretical Physics, Centre for
Mathematical Sciences, Wilberforce Road, University of Cambridge, Cambridge
CB3 0WA, United Kingdom}
\author{John~D.~Barrow}
\email[Email address: ]{J.D.Barrow@damtp.cam.ac.uk}
\affiliation{Department of Applied Mathematics and Theoretical Physics, Centre for
Mathematical Sciences, Wilberforce Road, University of Cambridge, Cambridge
CB3 0WA, United Kingdom}
\author{David~F.~Mota }
\email[Email address: ]{d.mota@thphys.uni-heidelberg.de}
\affiliation{Institute of Theoretical Physics, University of Heidelberg, 69120
Heidelberg, Germany}
\date{\today}

\begin{abstract}
We consider the cosmology where some function $f(G)$ of the
Gauss-Bonnet term $G$ is added to the gravitational action to
account for the late-time accelerating expansion of the universe.
The covariant and gauge invariant perturbation equations are
derived with a method which could also be applied to general
$f(R,R^{ab}R_{ab},R^{abcd}R_{abcd})$ gravitational theories. It is
pointed out that, despite their fourth-order character, such
$f(G)$ gravity models generally cannot reproduce arbitrary
background cosmic evolutions; for example, the standard $\Lambda
\mathrm{CDM}$ paradigm with $\Omega _{\mathrm{DE}}=0.76$ cannot be
realized in $f(G)$ gravity theories unless $f$ is a true
cosmological constant because it imposes exclusionary constraints
on the form of $f(G)$. We analyze the perturbation equations and
find that, as in $f(R)$ model, the stability of early-time
perturbation growth puts some constraints on the functional form
of $f(G)$, in this case $\partial ^{2}f/\partial G^{2}<0$.
Furthermore, the stability of small-scale perturbations also
requires that $f$ not deviate significantly from a constant. These
analyses are illustrated by numerically propagating the
perturbation equations with a specific model reproducing a
representative $\Lambda \mathrm{CDM}$ cosmic history. Our results
show how the $f(G)$ models are highly constrained by cosmological
data.
\end{abstract}

\maketitle












\section{Introduction}

\label{sect:Introduction}

The accumulating astronomical evidence for the present accelerating
expansion of the universe has stimulated many investigations into the nature
of the dark energy which might be responsible for this unexpected dynamics
(for a review see, \emph{e.g.}, \cite{DEReview}). Besides proposing to add
some kinds of exotic (and purely theoretical) matter species into the energy
budget of the universe, many investigators have also focused their
attentions on modifying general relativity (GR) on the largest scales so as
to introduce significant modifications at late times. One example is
provided by the family of $f(R)$ gravity models, which had also been
considered before the discovery of cosmic acceleration (see for example
Refs.~\cite{BOtt, BCot, Maeda1989}). In Refs.~\cite{Carroll2005, Easson2005}%
, the authors discuss a specific model where the correction to GR is a
polynomial function of the $R^{2},R^{ab}R_{ab}$ and $R^{abcd}R_{abcd}$
quadratic curvature invariants (here, $R,R_{ab}$ and $R_{abcd}$ are
respectively the Ricci scalar, Ricci tensor and Riemann tensor) and showed
that there exist late-time accelerating attractors in Friedmann cosmological
solutions to the theory. It is very interesting that when the Ricci scalar $R
$ in the Einstein-Hilbert action is replaced by some general functions of $R$
and $R^{ab}R_{ab}$, it becomes necessary to distinguish between two
different variational approaches in deriving the field equations. In the
metric approach, as in Refs.~\cite{Carroll2005, Easson2005}, the metric
components $g_{ab}$ are the only variational variables and the field
equations are generally of fourth-order, which makes the theories
phenomenologically richer but more stringently constrained in many cases.
Within the Palatini variational approach, on the other hand, we treat the
metric $g_{ab}$ and connection $\Gamma _{bc}^{a}$ as independent variables
and extremize the action with respect to both of them, and the resulting
field equations are second order and easier to solve. The Palatini $f(R)$
gravity is also proposed as an alternative to dark energy in a series of
works \cite{Vollick2003, Allemandi2004a, Allemandi2004b, Allemandi2005}.
There has since been growing interest in these modified gravity theories:
for the local tests of the Palatini and metric $f(R)$ gravity models see
\cite{PalatiniLGT, MetricLGT}, and for the cosmologies of these two classes
of models see \cite{PalatiniCT, MetricCT, Song2007, Li2007, Koivisto2006,
Li2006a, Li2006b}.

Both approaches to modifying gravity are far from problem-free. In the
metric $f(R)$ gravity models, the theory is conformally related to standard
GR plus a self-interacting scalar field \cite{BCot}, which generally
introduces extra forces inconsistent with solar system tests \cite{MetricLGT}%
. The Palatini approach, on the other hand, generally leads to a large (or
even negative) sound-speed-squared term in the growth equation of the matter
perturbations on small scales, and its predicted shapes of the cosmic
microwave background (CMB) and matter power-spectra deviate unacceptably
from those predicted in GR \cite{Koivisto2006, Li2006a, Li2006b}, and so
fail the most fundamental cosmological tests. Again, these examples
reiterate the difficulties encountered when trying to make general modified
gravity theories compatible with observations.

In this work we will focus on another form of modified gravity, the modified
Gauss-Bonnet theory, which is proposed and discussed in \cite{Nojiri2005,
NOG2006, Cognola2006a, Cognola2006b, Cognola2007} (see, for example, \cite%
{Noodsa2005, Carneu2006, Koimota2007} for a related model). In these models,
GR is modified by adding an arbitrary function $f(G)$ into the gravitational
action, where $G\equiv R^{2}-4R^{ab}R_{ab}+R^{abcd}R_{abcd}$ is the
Gauss-Bonnet invariant (which is a topological invariant in four
dimensions). Such correction is motivated by the effective low-energy
actions in string theory and is shown to be able to pass solar-system tests
even though it arises in the \emph{metric} variational approach.
Furthermore, refs. \cite{Nojiri2005, Cognola2006b} also demonstrate that
such models can produce late-time cosmic acceleration, as well as a
transition from deceleration to acceleration, or from a non-phantom phase to
a phantom phase. Here, we are interested in the perturbation dynamics in
such modified gravity theories, and the cosmology arising from them at
first-order in perturbation theory.

Our presentation is organized as follows. In Sec.~\ref{sect:Equations}, we
briefly introduce the model and present the perturbation equations of
general $f(G)$ model in covariant and gauge invariant (CGI) form. For this,
we generalize the effective energy-momentum tensor approach of deriving CGI
modified gravitational field equations in a way which could be applied to a
general $f(R,R^{ab}R_{ab},R^{abcd}R_{abcd})$ theory -- which includes the $%
f(G)$ and $f(R)$ models as specific cases. In Sec.~\ref{sect:Background} we
shall discuss the background evolution of $f(G)$ models and explain why
arbitrary cosmic histories cannot be realized with the $f(G)$ model. The
perturbation equations are then analyzed in Sec.~\ref{sect:Perturbation},
where we also evolve these equations numerically to investigate the $f(G)$
effects on the growth of linear perturbations. Our discussion and
conclusions are presented in Sec.~\ref{sect:Conclusion}. Throughout this
work our convention is chosen as $[\nabla _{a},\nabla _{b}]u^{c}=R_{ab\
d}^{\ \ c}u^{d},R_{ab}=R_{acb}^{\ \ \ c}$ where $a,b,\cdots $ run over $%
0,1,2,3$ and $c=\hslash =1$; the metric signature is $(+,-,-,-)$ and the
universe is assumed to be spatially flat and filled with photons, baryons,
cold dark matter (CDM) and three species of effectively massless neutrinos.

\section{Field Equations in Modified Gauss-Bonnet Gravity}

\label{sect:Equations}

In this section we briefly introduce the main ingredients of $f(G)$ gravity
and derive the general perturbation equations that govern the dynamics of
small inhomogeneities in the cosmological models that arise in this theory.

\subsection{The Generalized Einstein Equations}

Our starting point for $f(G)$ gravity is the modified Einstein-Hilbert
action,
\begin{eqnarray}  \label{eq:1}
S &=& \int d^{4}x\sqrt{-g}\left[ \frac{R+f(G)}{2\kappa} + \mathcal{L}_{m}%
\right],
\end{eqnarray}
in which $\kappa =8\pi G_{\mathrm{N}},$ with $G_{\mathrm{N}}$ being the
gravitational constant and $R=R(g_{ab})$ is the Ricci scalar. Varying this
action with respect to the metric $g_{ab}$ gives the modified Einstein
equations
\begin{eqnarray}  \label{eq:2}
&&R_{ab}-\frac{1}{2}g_{ab}R  \nonumber \\
&=& \kappa T_{ab}^{f}+\frac{1}{2}g_{ab}f-2FRR_{ab}+4FR_{a}^{c}R_{bc}
\nonumber \\
&&-2FR_{acde}R_{b}^{\ cde}-4FR_{acdb}R^{cd}+2R\nabla _{a}\nabla _{b}F
\nonumber \\
&&-2Rg_{ab}\nabla ^{2}F-4R_{a}^{c}\nabla _{b}\nabla _{c}F-4R_{b}^{c}\nabla
_{a}\nabla _{c}F  \nonumber \\
&&+4R_{ab}\nabla ^{2}F+4g_{ab}R^{cd}\nabla _{c}\nabla _{d}F-4R_{acbd}\nabla
^{c}\nabla ^{d}F,\ \ \
\end{eqnarray}%
where $F=F(G)\equiv \partial f(G)/\partial G$ (notice that, unlike in the $%
f(R)$ models, $F$ here is not dimensionless) and $T_{ab}^{f}$ is the
energy-momentum tensor of the fluid matter (photons, baryons, cold dark
matter, and light neutrinos). The trace of Eq.~(\ref{eq:2}) reads
\begin{eqnarray}  \label{eq:3}
-R &=&\kappa (\rho ^{f}-3p^{f})+2f-2FG  \nonumber \\
&&-2R\Box F+4R_{ab}\nabla ^{a}\nabla ^{b}F,
\end{eqnarray}%
with $T_{a}^{a}\equiv T=\rho ^{f}-3p^{f}$. We see that the curvature-related
quantities ($R,G,F,\cdots $) are determined by the energy-momentum tensor of
the fluid matter through a complicated dynamical equation Eq.~(\ref{eq:3})
and thus the modification to the GR field equations can be understood as a
change in the way that the spacetime curvature, and thus the Ricci tensor $%
R_{ab}$, responds to the distribution of matter.

\subsection{The Perturbation Equations in General Relativity}

The CGI perturbation equations in general theories of $f(G)$ gravity are
derived in this section using the method of $3+1$ decomposition \cite%
{Ellis1989, Ellis1998, Challinor1999, Lewis2000}. Furthermore, we shall
adopt the effective energy-momentum tensor approach \cite{Hwang1990, Li2007}%
, which treats the modifications on the right hand side of Eq.~(\ref{eq:2})
as an effective energy-momentum tensor. However, since the modification
generally involves terms nonlinear in $R_{ab},R_{abcd}$, we should express
these terms appropriately. This will be done in more detail below, but now
let us briefly review the main ingredients of $3+1$ decomposition and their
application to standard general relativity for ease of later reference (as
in GR, there is only fluid matter so we shall neglect the superscript $^{f}$
in this subsection).

The main idea of $3+1$ decomposition is to make spacetime splits of physical
quantities with respect to the 4-velocity $u^{a}$ of an observer. The
projection tensor $h_{ab}$ is defined as $h_{ab}=g_{ab}-u_{a}u_{b}$ and can
be used to obtain covariant tensors perpendicular to $u$. For example, the
covariant spatial derivative $\hat{\nabla}$ of a tensor field $T_{d\cdot
\cdot \cdot e}^{b\cdot \cdot \cdot c}$ is defined as
\begin{eqnarray}  \label{eq:4}
\hat{\nabla}^{a}T_{d\cdot \cdot \cdot e}^{b\cdot \cdot \cdot c}\equiv
h_{i}^{a}h_{j}^{b}\cdot \cdot \cdot \ h_{k}^{c}h_{d}^{r}\cdot \cdot \cdot \
h_{e}^{s}\nabla ^{i}T_{r\cdot \cdot \cdot s}^{j\cdot \cdot \cdot k}.
\end{eqnarray}%
The energy-momentum tensor and covariant derivative of the 4-velocity are
decomposed respectively as
\begin{eqnarray}
\label{eq:5}
T_{ab} &=&\pi _{ab}+2q_{(a}u_{b)}+\rho u_{a}u_{b}-ph_{ab}, \\
 \label{eq:6} \nabla _{a}u_{b} &=&\sigma _{ab}+\varpi _{ab}+\frac{1}{3}\theta
h_{ab}+u_{a}A_{b}.
\end{eqnarray}%
In the above, $\pi _{ab}$ is the projected symmetric trace-free (PSTF)
anisotropic stress, $q_{a}$ the vector heat flux vector, $p$ the isotropic
pressure, $\sigma _{ab}$ the PSTF shear tensor, $\varpi _{ab}=\hat{\nabla}%
_{[a}u_{b]}$, the vorticity, $\theta =\nabla ^{c}u_{c}=3\dot{a}/a$ ($a$ is
the mean expansion scale factor) the expansion scalar, and $A_{b}=\dot{u}%
_{b} $ the acceleration; the overdot denotes time derivative expressed as $%
\dot{\phi}=u^{a}\nabla _{a}\phi $, brackets mean antisymmetrisation, and
parentheses symmetrization. The normalization is chosen to be $u^{a}u_{a}=1$%
. The quantities $\pi _{ab},q_{a},\rho ,p$ are referred to as \emph{dynamical%
} quantities and $\sigma _{ab},\varpi _{ab},\theta ,A_{a}$ as \emph{%
kinematical} quantities. Note that the dynamical quantities can be obtained
from the energy-momentum tensor $T_{ab}$ through the relations
\begin{eqnarray}  \label{eq:7}
\rho &=&T_{ab}u^{a}u^{b},  \nonumber \\
p &=&-\frac{1}{3}h^{ab}T_{ab},  \nonumber \\
q_{a} &=&h_{a}^{d}u^{c}T_{cd},  \nonumber \\
\pi _{ab} &=&h_{a}^{c}h_{b}^{d}T_{cd}+ph_{ab}.
\end{eqnarray}

Decomposing the Riemann tensor and making use the Einstein equations, we
obtain, after linearization, five constraint equations \cite{Challinor1999,
Lewis2000}:
\begin{eqnarray}
\label{eq:8} 0 &=&\hat{\nabla}^{c}(\varepsilon _{\ \ cd}^{ab}u^{d}\varpi _{ab}); \\
\label{eq:9} \kappa q_{a} &=& -\frac{2\hat{\nabla}_{a}\theta}{3} +
\hat{\nabla}^{b}\sigma_{ab}+\hat{\nabla}^{b}\varpi _{ab};\ \ \  \\
\label{eq:10} \mathcal{B}_{ab} &=&\left[ \hat{\nabla}^{c}\sigma
_{d(a}+\hat{\nabla}^{c}\varpi _{d(a}\right] \varepsilon _{b)ec}^{\ \ \ \ d}u^{e}; \\
\label{eq:11} \hat{\nabla}^{b}\mathcal{E}_{ab} &=&
\frac{1}{2}\kappa \left[\hat{\nabla}^{b}\pi_{ab}+\frac{2}{3}\theta
q_{a}+\frac{2}{3}\hat{\nabla}_{a}\rho \right];\\
\label{eq:12} \hat{\nabla}^{b}\mathcal{B}_{ab}
&=&\frac{1}{2}\kappa \left[\hat{\nabla}_{c}q_{d}+(\rho +p)\varpi
_{cd}\right] \varepsilon _{ab}^{\ \ cd}u^{b},
\end{eqnarray}%
and five propagation equations,
\begin{eqnarray}
\label{eq:13} \dot{\theta}+\frac{1}{3}\theta ^{2}-\hat{\nabla}^{a}A_{a}+\frac{\kappa }{2}%
(\rho +3p) &=& 0; \\
\label{eq:14} \dot{\sigma}_{ab}+\frac{2}{3}\theta \sigma
_{ab}-\hat{\nabla}_{\langle
a}A_{b\rangle }+\mathcal{E}_{ab}+\frac{1}{2}\kappa \pi _{ab} &=& 0; \\
\label{eq:15} \dot{\varpi}+\frac{2}{3}\theta \varpi -\hat{\nabla}_{[a}A_{b]} &=& 0; \\
\label{eq:16} \frac{1}{2}\kappa \left[\dot{\pi}_{ab} +
\frac{1}{3}\theta\pi_{ab}\right] - \frac{1}{2}\kappa \left[(\rho
+p)\sigma_{ab}+\hat{\nabla}_{\langle
a}q_{b\rangle}\right]  \nonumber \\
-\left[ \dot{\mathcal{E}}_{ab}+\theta \mathcal{E}_{ab}-\hat{\nabla}^{c}%
\mathcal{B}_{d(a}\varepsilon _{b)ec}^{\ \ \ \ d}u^{e}\right] &=& 0;\ \ \ \
\\
\label{eq:17} \dot{\mathcal{B}}_{ab}+\theta \mathcal{B}_{ab}+\hat{\nabla}^{c}\mathcal{E}%
_{d(a}\varepsilon _{b)ec}^{\ \ \ \ d}u^{e}  \nonumber \\
+\frac{\kappa }{2}\hat{\nabla}^{c}\mathcal{\pi }_{d(a}\varepsilon _{b)ec}^{\
\ \ \ d}u^{e} &=& 0.
\end{eqnarray}%
Here, $\varepsilon _{abcd}$ is the covariant permutation tensor, $\mathcal{E}%
_{ab}$ and $\mathcal{B}_{ab}$ are respectively the electric and magnetic
parts of the Weyl tensor $\mathcal{W}_{abcd}$, defined by $\mathcal{E}%
_{ab}=u^{c}u^{d}\mathcal{W}_{acbd}$ and $\mathcal{B}_{ab}=-\frac{1}{2}%
u^{c}u^{d}\varepsilon _{ac}^{\ \ ef}\mathcal{W}_{efbd}$. The angle bracket
means taking the trace-free part of a quantity.

Besides the above equations, it is useful to express the projected Ricci
scalar $\hat{R}$ into the hypersurfaces orthogonal to $u^{a}$ as
\begin{eqnarray}  \label{eq:18}
\hat{R} &\doteq& 2\kappa\rho - \frac{2}{3}\theta ^{2}.
\end{eqnarray}%
The spatial derivative of the projected Ricci scalar, $\eta _{a}\equiv \frac{%
1}{2}a\hat{\nabla}_{a}\hat{R}$, is then given as
\begin{eqnarray}  \label{eq:19}
\eta _{a} &=& \kappa \hat{\nabla}_{a}\rho - \frac{2a}{3}\theta\hat{\nabla}%
_{a}\theta,
\end{eqnarray}%
and its propagation equation by
\begin{eqnarray}  \label{eq:20}
\dot{\eta}_{a}+\frac{2\theta }{3}\eta _{a} &=& -\frac{2}{3}\theta a\hat{%
\nabla}_{a}\hat{\nabla}\cdot A - a\kappa\hat{\nabla}_{a}\hat{\nabla}\cdot q.
\end{eqnarray}

Finally, there are the conservation equations for the energy-momentum
tensor:
\begin{eqnarray}  \label{eq:21}
\dot{\rho}+(\rho +p)\theta +\hat{\nabla}^{a}q_{a} &=&0, \\
\label{eq:22} \dot{q}_{a}+\frac{4}{3}\theta q_{a}+(\rho +p)A_{a}-\hat{\nabla}_{a}p+\hat{%
\nabla}^{b}\pi _{ab} &=&0.
\end{eqnarray}

As we are considering a spatially-flat universe, the spatial curvature must
vanish on large scales and so $\hat{R}=0$. Thus, from Eq.~(\ref{eq:18}), we
obtain
\begin{eqnarray}  \label{eq:23}
\frac{1}{3}\theta ^{2}=\kappa\rho.
\end{eqnarray}
This is the Friedmann equation in standard general relativity, and the other
background equations (the Raychaudhuri equation and the energy-conservation
equation) are obtained by taking the zero-order parts of Eqs.~(\ref{eq:13}, %
\ref{eq:21}), as
\begin{eqnarray}
\label{eq:24} \dot{\theta}+\frac{1}{3}\theta ^{2}+\frac{\kappa }{2}(\rho +3p) &=& 0, \\
\label{eq:25} \dot{\rho}+(\rho +p)\theta &=&0.
\end{eqnarray}

In what follows, we will only consider scalar modes of perturbations, for
which the vorticity $\varpi _{ab}$ and magnetic part of Weyl tensor $%
\mathcal{B}_{ab}$ are at most of second order \cite{Challinor1999, Lewis2000}
and will be neglected from our first-order analysis.

\subsection{The Perturbation Equations in $f(G)$ Gravity}

In the effective energy-momentum tensor approach, the field equations Eqs.~(%
\ref{eq:8} - \ref{eq:25}) listed above preserve their forms, but
the dynamical quantities $\rho ,p,q_{a},\pi _{ab}$ appearing there
should be replaced with the effective total ones $\rho
^{\mathrm{tot}}=\rho ^{f}+\rho ^{\mathrm{G}},
p^{\mathrm{tot}} = p^{f}+p^{\mathrm{G}}, q_{a}^{\mathrm{tot}%
}=q_{a}^{f}+q_{a}^{\mathrm{G}}, \pi_{ab}^{\mathrm{tot}}=\pi_{ab}^{f}+%
\pi_{ab}^{\mathrm{G}}$, in which a superscript $^{\mathrm{G}}$ means the
contribution from the Gauss-Bonnet correction.

Writing the modified Einstein equations, Eq.~(\ref{eq:2}) in the following
form,
\begin{eqnarray}  \label{eq:26}
R_{ab}-\frac{1}{2}g_{ab}R &=&\kappa T_{ab}^{\mathrm{tot}}  \nonumber \\
&=& \kappa T_{ab}^{f}+\kappa T_{ab}^{\mathrm{G}},
\end{eqnarray}%
one can easily identify
\begin{eqnarray}
\kappa T_{ab}^{\mathrm{G}} &=&\frac{1}{2}g_{ab}f-2FRR_{ab}+4FR_{a}^{c}R_{bc}
\nonumber \\
&&-2FR_{acde}R_{b}^{\ cde}-4FR_{acdb}R^{cd}+2R\nabla _{a}\nabla _{b}F
\nonumber \\
&&-2Rg_{ab}\nabla ^{2}F-4R_{a}^{c}\nabla _{b}\nabla _{c}F-4R_{b}^{c}\nabla
_{a}\nabla _{c}F  \nonumber \\
&&+4R_{ab}\nabla ^{2}F+4g_{ab}R^{cd}\nabla _{c}\nabla _{d}F-4R_{acbd}\nabla
^{c}\nabla ^{d}F  \nonumber
\end{eqnarray}%
and Eq.~(\ref{eq:7}) can be used to calculate $\rho ^{\mathrm{G}},p^{\mathrm{%
G}},q_{a}^{\mathrm{G}},\pi _{ab}^{\mathrm{G}}$. In order to do this we need
the explicit expressions for $R_{ab}$ and $R_{abcd}$, which could be
obtained in terms of either (effective total) dynamical quantities or
kinematical quantities, or a mixture of the two.

To express $R_{ab}$ and $R_{abcd}$ explicitly, now decompose the symmetric
Ricci tensor $R_{ab}$ in the following general way,
\begin{eqnarray}  \label{eq:27}
R_{ab} &=& \Delta u_{a}u_{b}+\Xi h_{ab}+2u_{(a}\Upsilon _{b)}+\Sigma _{ab},
\end{eqnarray}%
then Eq.~(\ref{eq:26}) gives
\begin{eqnarray}  \label{eq:28}
\Delta &=&\frac{1}{2}\kappa (\rho ^{\mathrm{tot}}+3p^{\mathrm{tot}})
\nonumber \\
&=&-\left[ \dot{\theta}+\frac{1}{3}\theta ^{2}-\hat{\nabla}^{a}A_{a}\right] ;
\nonumber \\
\Xi &=&-\frac{1}{2}\kappa (\rho ^{\mathrm{tot}}-p^{\mathrm{tot}})  \nonumber
\\
&=&-\frac{1}{3}\left[ \dot{\theta}+\theta ^{2}+\hat{R}-\hat{\nabla}^{a}A_{a}%
\right] ;  \nonumber \\
\Upsilon _{a} &=&\kappa q_{a}^{\mathrm{tot}}  \nonumber \\
&=&-\frac{2\hat{\nabla}_{a}\theta }{3}+\hat{\nabla}^{b}\sigma _{ab}+\hat{%
\nabla}^{b}\varpi _{ab};  \nonumber \\
\Sigma _{ab} &=&\kappa \pi _{ab}^{\mathrm{tot}}  \nonumber \\
&=&-2\left[ \dot{\sigma}_{ab}+\frac{2}{3}\theta \sigma _{ab}-\hat{\nabla}%
_{\langle a}A_{b\rangle }+\mathcal{E}_{ab}\right]
\end{eqnarray}%
in which we have used Eqs.~(\ref{eq:9}, \ref{eq:13}, \ref{eq:14}, \ref{eq:18}%
). Notice that the first lines are expressed in terms of total dynamical
quantities and the second lines of kinematical quantities. For those terms
involving $R_{abcd},$ we shall use the decomposition of Riemann tensor
extensively (keeping in mind that $u^{c}u^{d}\mathcal{W}_{acbd}=\mathcal{E}%
_{ab}$):
\begin{eqnarray}  \label{eq:29}
R_{abcd} &=&\frac{1}{2}(g_{ac}R_{bd}+g_{bd}R_{ac}-g_{ad}R_{bc}-g_{bc}R_{ad})
\nonumber \\
&&+\mathcal{W}_{abcd}-\frac{1}{6}R(g_{ac}g_{bd}-g_{ad}g_{bc}).
\end{eqnarray}%
For example, it is easy to show that, up to first order,
\begin{eqnarray}  \label{eq:30}
&& R_{a}^{\ cde}R_{bcde}  \nonumber \\
&=&-\frac{4}{3}\dot{\theta}\mathcal{E}_{ab}+\frac{1}{2}g_{ab}R^{cd}R_{cd}+%
\frac{1}{3}RR_{ab}-\frac{1}{6}g_{ab}R^{2},
\end{eqnarray}%
so that
\begin{eqnarray}  \label{eq:31}
R^{abcd}R_{abcd} &=&2R^{ab}R_{ab}-\frac{1}{3}R^{2};  \nonumber \\
G &=&\frac{2}{3}R^{2}-2R^{ab}R_{ab}
\end{eqnarray}%
where
\begin{eqnarray}  \label{eq:32}
R &=& -2\dot{\theta}-\frac{4}{3}\theta ^{2}+2\hat{\nabla}^{a}A_{a}-\hat{R};
\nonumber \\
R^{ab}R_{ab} &=&\frac{4}{3}\left[ \dot{\theta}^{2}+\dot{\theta}\theta ^{2}+%
\frac{1}{3}\theta ^{4}\right]  \nonumber \\
&&+\frac{2}{3}(\dot{\theta}+\theta ^{2})\hat{R}-\frac{8}{3}\left[ \dot{\theta%
}+\frac{1}{2}\theta ^{2}\right] \hat{\nabla}^{a}A_{a}.
\end{eqnarray}

With these useful relations and some calculation, the contribution to the
energy-momentum tensor from the Gauss-Bonnet correction term can be
identified as
\begin{eqnarray}
\label{eq:33} \kappa \rho ^{\mathrm{G}} &=&\frac{1}{2}(f-FG)+\frac{2}{3}(\Delta -3\Xi )%
\dot{F}\theta  \nonumber \\
&& + \frac{2}{3}(\Delta -3\Xi )\hat{\nabla}^{2}F; \\
\label{eq:34} -\kappa p^{\mathrm{G}} &=&\frac{1}{2}(f-FG)+\frac{2}{3}(\Delta -3\Xi )\ddot{F%
}  \nonumber \\
&&-\frac{8}{9}\Delta (\theta \dot{F}+\hat{\nabla}^{2}F); \\
\label{eq:35} \kappa q_{a}^{\mathrm{G}} &=&-\frac{2}{3}(\Delta -3\Xi )\left( \hat{\nabla}%
_{a}\dot{F}-\frac{1}{3}\theta \hat{\nabla}_{a}F\right) +\frac{4}{3}\dot{F}%
\theta \Upsilon _{a};\ \ \  \\
\label{eq:36} \kappa \pi _{ab}^{\mathrm{G}} &=&\frac{4}{3}\Delta
\left( \dot{F}\sigma _{ab}+\hat{\nabla}_{\langle
a}\hat{\nabla}_{b\rangle }F\right) +2\left(
\ddot{F}+\frac{1}{3}\theta \dot{F}\right) \Sigma _{ab}  \nonumber \\
&&-4\ddot{F}\mathcal{E}_{ab}+\frac{4}{3}\theta \dot{F}\mathcal{E}_{ab}.
\end{eqnarray}%
Here, we want to make some comments about these equations. First,
if $f$ is constant, then $F$ and its derivatives vanish, so that
$\kappa \rho^{\mathrm{G}}=-\kappa p^{\mathrm{G}}=\frac{f}{2}$ and
$\kappa q_{a}^{\mathrm{G}}=\kappa \pi _{ab}^{\mathrm{G}}=0$, and
thus we have the $\Lambda \mathrm{CDM}$ limit. Second, it is not
difficult to check that the above quantities satisfy the
independent energy-momentum conservation equations
\begin{eqnarray}
\dot{\rho}^{\mathrm{G}}+(\rho ^{\mathrm{G}}+p^{\mathrm{G}})\theta +\hat{%
\nabla}^{a}q_{a}^{\mathrm{G}} &=&0,  \nonumber \\
\dot{q}_{a}^{\mathrm{G}}+\frac{4}{3}\theta q_{a}^{\mathrm{G}}+(\rho ^{%
\mathrm{G}}+p^{\mathrm{G}})A_{a}-\hat{\nabla}_{a}p^{\mathrm{G}}+\hat{\nabla}%
^{b}\pi _{ab}^{\mathrm{G}} &=&0.  \nonumber
\end{eqnarray}%
This is a result of the energy-momentum conservation in fluid matter and the
contracted Bianchi identity. Thirdly, it would be convenient to use $%
\Upsilon _{a}=\kappa (q_{a}^{f}+q_{a}^{\mathrm{G}})$ and $\Sigma
_{ab}=\kappa (\pi _{ab}^{f}+\pi _{ab}^{\mathrm{G}})$ (see Eq.~(\ref{eq:28}))
to rewrite Eqs.~(\ref{eq:35}, \ref{eq:36}) so that $\kappa q_{a}^{\mathrm{G}%
},\kappa \pi _{ab}^{\mathrm{G}}$ are expressed respectively in terms of $%
\kappa q_{a}^{f},\kappa \pi _{ab}^{f}$ and other quantities. This is what we
do in the numerical calculation. Fourthly, it is interesting to note that
there is no $\hat{\nabla}_{a}\dot{\theta}$ term in $\kappa \hat{\nabla}%
_{a}\rho ^{\mathrm{G}}$ up to first order because $\Delta -3\Xi =\frac{2}{3}%
\theta ^{2}+\hat{R}$ and $\hat{\nabla}_{a}(f-FG)=-G\hat{\nabla}_{a}F$; this
is positive because otherwise Eq.~(\ref{eq:19}) will no longer be an
algebraic equation for $\hat{\nabla}_{a}\theta $. For similar reasons Eq.~(%
\ref{eq:20}) remains a first-order differential equation for $\eta _{a}$ for
the present model. These simplifications also occur in the $f(R)$ gravity
models but not in general $f(R,R^{ab}R_{ab},R^{abcd}R_{abcd})$ theories. In
the later case, the method we use here to derive the CGI perturbation
equations still applies and the perturbation equations will become even
higher order and more complicated (specifically, some of the perturbation
equations above will become propagation equations for $q_{a}^{\mathrm{MG}}$
and $\pi _{ab}^{\mathrm{MG}},$ where $^{\mathrm{MG}}$ denotes general
modified gravity theory \cite{Li2007b, Comment1}). Finally, we can see that
the quantity $F$ here appears only to (at least directly) influence
background evolutions, and with the background fixed it is its derivative
which determines the perturbation evolutions.

\section{The Background Evolution in $f(G)$ Models}

\label{sect:Background}

In this section we discuss the background evolution in general $f(G)$
gravity models. Recall that in $f(R)$ gravity theories the fourth-order
nature of the Friedmann equation allows enough freedom for the model to
reproduce an arbitrary background cosmic evolution. Since the field
equations are also fourth order in the $f(G)$ models, one might think that
they could also describe arbitrarily parameterized background histories.
However, this is not the case, as we shall see below. This is because $\dot{G%
}$ must change its sign in the recent past in many fixed-background models.

For the background evolution, we use the Friedmann equation Eq.~(\ref{eq:23}%
) with the dark energy density given by Eq.~(\ref{eq:33})
\begin{eqnarray}  \label{eq:37}
\kappa\rho _{\mathrm{DE}} &=& \frac{1}{2}f-\frac{4}{9}\theta ^{2}\left( \dot{%
\theta}+\frac{1}{3}\theta ^{2}\right) F+\frac{4}{9}\theta ^{3}\dot{F}
\end{eqnarray}%
where we have kept terms only up to zero order and used the expression $G =
\frac{8}{9}\theta ^{2}\dot{\theta}+\frac{8}{27}\theta ^{4}$, which is
obtained from Eqs.~(\ref{eq:27}, \ref{eq:28}, \ref{eq:31}). Following \cite%
{Song2007}, we define the following dimensionless quantities (here $H=\theta
/3$ is the Hubble rate and $H_{0}$ is its present-day value)
\begin{eqnarray}  \label{eq:38}
E &\equiv& \frac{H^{2}}{H_{0}^{2}},  \nonumber  \label{eq:38} \\
y &\equiv &\frac{f}{H_{0}^{2}},
\end{eqnarray}%
in terms of which $G,$ and so Eq.~(\ref{eq:37}) can be written as
\begin{eqnarray}  \label{eq:39}
G &=& 12H_{0}^{4}E(E^{\ast}+2E),
\end{eqnarray}%
with
\begin{eqnarray}  \label{eq:40}
&& y^{\ast \ast }-\left( \frac{G}{24E^{2}}+\frac{G^{\ast \ast }}{G^{\ast }}%
\right) y^{\ast }+\frac{G^{\ast }}{24E^{2}}y  \nonumber \\
&=& \frac{G^{\ast }}{4E^{2}}\Omega _{\mathrm{DE}}\exp \left[ -3(1+w)N\right]
,
\end{eqnarray}%
where a star denotes the derivative with respect to $N=\log (a)$, $\Omega _{%
\mathrm{DE}}\equiv \kappa \rho _{\mathrm{DE}}/3H_{0}^{2}$ is the dark energy
fractional energy density today, and $w=\mathrm{const}.$ is the usual
dark-energy equation of state (EOS) parameter. Note that by writing in this
way we have chosen to parameterize the background expansion to be the same
as the dynamical dark-energy model $(w\neq -1)$ or the $\Lambda \mathrm{CDM}$
paradigm ($w=-1$).

In the following, we shall assume $w=-1$ for simplicity and the calculations
for general $w$ could be done similarly. In this case we have
\begin{eqnarray}  \label{eq:41}
E &=& \Omega _{\mathrm{DE}}+\Omega _{m}\exp (-3N)+\Omega _{r}\exp (-4N),
\end{eqnarray}%
where $\Omega _{m},\Omega _{ra}$ are respectively the fractional densities
for non-relativistic and relativistic matter species. Deep into the
radiation-dominated era we have $E\doteq \Omega _{r}\exp (-4N),$ and thus
Eq.~(\ref{eq:40}) reduces to
\begin{eqnarray}  \label{eq:42}
y^{\ast\ast}+9y^{\ast}+8y &=& 48\Omega _{\mathrm{DE}},
\end{eqnarray}%
whose solution is
\begin{eqnarray}  \label{eq:43}
y(N) &=& A\exp (-N)+B\exp (-8N)+6\Omega _{\mathrm{DE}}.
\end{eqnarray}%
In this work we shall require
\begin{eqnarray}
\lim_{|G,R|\rightarrow \infty }\left\vert \frac{f(G)}{R}\right\vert
\rightarrow 0  \nonumber
\end{eqnarray}
so that $B$ could be set to zero in Eq.~(\ref{eq:43}). To obtain the
background evolution numerically, we start deep in the radiation-dominated
era (\emph{e.g.,} at $a=10^{-6}$) and take the radiation-dominated solution,
Eq.~(\ref{eq:43}), to be the initial condition. The solutions to Eq.~(\ref%
{eq:40}) are then characterized by a single parameter $A$. Different values
of $A$ give the same background history, but in general lead to different
evolutions for the perturbations, as we shall see below. Note that $A=0$
describes the standard $\Lambda \mathrm{CDM}$ paradigm.

Up to this point the procedure is quite similar to that in metric $f(R)$
gravity models. However, as was claimed above, the $f(G)$ gravity model
cannot be used to reproduce arbitrarily parameterized background expansion
histories. To explain why, we shall again take the $\Lambda \mathrm{CDM}$
background as an example and adopt the value $\Omega _{\mathrm{DE}}=0.76$,
as suggested by the Wilkinson Microwave Anisotropy Probe (WMAP) three-year
data \cite{WMAP3}. Then it will be easy to find that $G^{\ast }$ changes its
sign (from $+$ to $-$) at $N_{0}\approx -0.153$, which means that $G$
increases (decreases) when $N<(>)N_{0}$. As a result, with the match to $%
\Lambda \mathrm{CDM}$ at $N<N_{0}$ the function $f(G)$ with \emph{all
possible values} of $G$ has been determined, and there will generally be no
freedom left to fix the evolution to $\Lambda \mathrm{CDM}$ at $N>N_{0}$ as
well (of course, if $f(G)$ is a real cosmological constant then $\Lambda
\mathrm{CDM}$ will also be reproduced at $N>N_{0}$, but in general
reproducing $\Lambda \mathrm{CDM}$ on \emph{both} sides of $N_{0}$ is far
too strong a requirement to be satisfied). So, what we may conclude is that
the $f(G)$ model can mimic a $\Lambda \mathrm{CDM}$ universe up to $N_{0}$,
after which the evolution might be governed by the already-determined $f(G)$%
. However, note that $N_{0}\approx -0.153$ corresponds to a critical
redshift $z_{0}\approx 0.166$, so the transition from $\Lambda \mathrm{CDM}$
phase to a non-$\Lambda \mathrm{CDM}$ phase occurs quite late \cite{Comment2}%
.

\begin{figure}[tbp]
\centering
\includegraphics[scale=0.9] {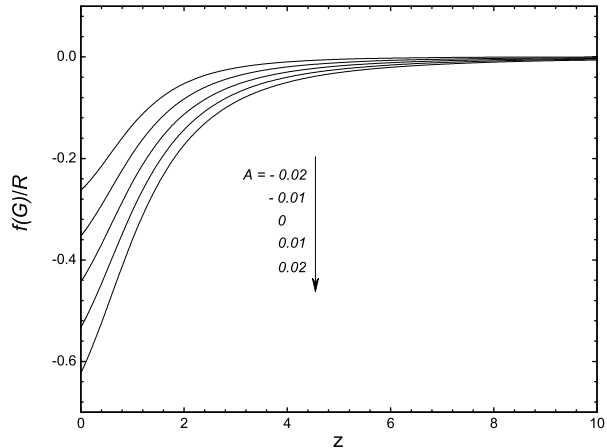}
\caption{The evolution versus redshift of $f(G)/R$ for the $f(G)$ gravity
models fixed to match a $\Lambda\mathrm{CDM}$ cosmic expansion history ($%
\Omega_{\mathrm{DE}} = 0.66$). The curves from top to bottom are
characterized by $A = -0.02, -0.01, 0\ (\Lambda\mathrm{CDM}), 0.01, 0.02$
respectively. Notice that with our convention $R < 0$ and $f > 0$.}
\label{fig:Figure1}
\end{figure}

\begin{figure}[tbp]
\centering
\includegraphics[scale=0.9] {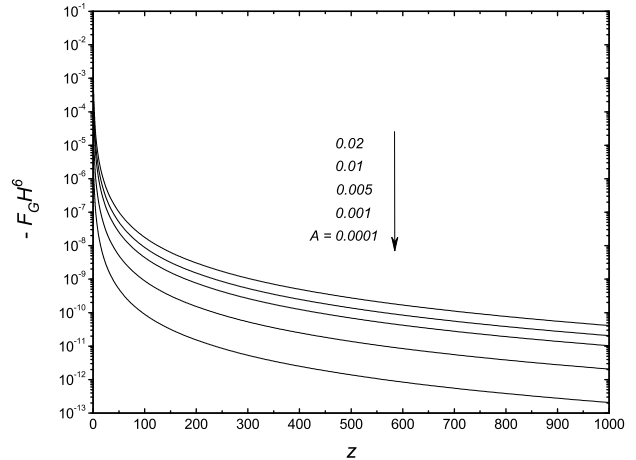}
\caption{The evolution versus redshift of the dimensionless quantity $%
-F_{G}H^{6}$ for the $f(G)$ gravity models fixed to match a $\Lambda\mathrm{%
CDM}$ cosmic expansion history ($\Omega_{\mathrm{DE}} = 0.66$) which are
stable in early time perturbation growths ($F_{G} < 0$). The curves from
bottom to top are characterized by $A = 0.0001, 0.001, 0.005, 0.01, 0.02$
respectively. Note that a true cosmological constant has $F_{G} = 0$.}
\label{fig:Figure2}
\end{figure}

In the above, we have analyzed for a $\Lambda \mathrm{CDM}$ ($w=-1$)
background. Nonetheless we can expect that similar situations exist for
general dynamical ($w\neq -1$) dark-energy backgrounds which are
characterized by a late-time transition from $G^{\ast }>0$ to $G^{\ast }<0$.
The investigation of background evolutions in general modified Gauss-Bonnet
models is an interesting topic which is unfortunately beyond the scope of
the present work. Here, we solve the perturbed equations numerically and
investigate the effects of the $f(G)$ modifications to GR on the growth of
linear perturbations. For simplicity we shall adopt a slightly unrealistic $%
\Lambda \mathrm{CDM}$ cosmic history which is described by $\Omega _{\mathrm{%
DE}}=0.66$ (in this case the transition from $G^{\ast }>0$ to $G^{\ast } < 0$
has not taken place yet).

In Fig.~\ref{fig:Figure1}, we show the redshift evolutions of $f(G)$
compared with $R$. Each curve here is characterized by a specific value of
the coefficient $A$, which can be either positive, negative or 0. Note that
although all these curves lead to the same $\Lambda \mathrm{CDM}$ background
evolution, not all of them are cosmologically viable. As will be discussed
below, the stability of early-time perturbation growth requires $F_{G}<0$
(where $F_{G}\equiv \partial F/\partial G$) and that of small-scale
perturbation growth furthermore requires $|F_{G}|H^{6}\ll 1$. The condition $%
F_{G}<0$ is found to correspond to the $A>0$ subclass of the solutions to
Eq.~(\ref{eq:40}). The evolution of $|F_{G}|H^{6}=-F_{G}H^{6}$ (which is
dimensionless) in this subclass are shown in Fig.~\ref{fig:Figure2}. It is
obvious that $|F_{G}|H^{6}$ is a rapidly increasing quantity with respect to
time whose magnitude is much smaller than 1 at early times.

\section{The Perturbation Evolution in $f(G)$ Models}

\label{sect:Perturbation}

Now we turn to the evolution of perturbation equations in the $f(G)$ model.
The equations presented in Sec.~\ref{sect:Equations} are not yet closed and
from Eqs.~(\ref{eq:33} - \ref{eq:36}) we see that an evolution equation for $%
\hat{\nabla}_{a}F$ is also needed. To obtain this, let us look at the trace
equation, Eq.~(\ref{eq:3}) (which can also be obtained from Eq.~(19) and the
spatial derivative of Eq.~(13)). Taking its spatial covariant derivative,
after some manipulations we obtain the following evolution equation
\begin{eqnarray}  \label{eq:44}
0 = \ddot{\epsilon}+\left( \theta +\frac{2\dot{\theta}}{\theta }\right) \dot{%
\epsilon}-S  \nonumber \\
+ \left[ \left( 1+\frac{2\dot{\theta}}{\theta ^{2}}\right) \frac{k^{2}}{a^{2}%
}-\frac{4}{3}\left( \dot{\theta}+\frac{1}{3}\theta ^{2}\right) -\frac{27}{16}%
\frac{1-\frac{4}{3}\dot{F}\theta }{\theta ^{4}F_{G}}\right]\epsilon\ \ \
\end{eqnarray}%
where $\epsilon $ is the harmonic expansion coefficient of $\hat{\nabla}%
_{a}F,$ as
\begin{eqnarray}  \label{eq:45}
\hat{\nabla}_{a}F &=& \sum_{k}\frac{k}{a}\epsilon Q_{a}^{k},
\end{eqnarray}%
and $Q_{a}^{k}=\frac{a}{k}\hat{\nabla}_{a}Q^{k}$ and $Q^{k}$ are the
zero-order eigenvalues of the comoving Laplacian $a^{2}\hat{\nabla}^{2}$ ($%
a^{2}\hat{\nabla}^{2}Q^{k}=k^{2}Q^{k}$), $F_{G}=\partial F/\partial
G=\partial ^{2}f/\partial G^{2}$, and the source function $S$ is given by
\begin{eqnarray}  \label{eq:46}
S &=&-\frac{3\kappa }{4\theta ^{2}}(\mathcal{X}^{f}-3\mathcal{X}%
^{pf})-3\left( 1-\frac{4}{3}\dot{F}\theta \right) \frac{\dot{\theta}}{\theta
^{3}}\frac{k}{a}\mathcal{Z}  \nonumber \\
&&-\frac{9}{2}\left( 1-\frac{4}{3}\dot{F}\theta \right) \frac{\dot{\theta}}{%
\theta ^{4}}\frac{k^{2}}{a^{2}}\eta -3\left( \frac{\ddot{F}}{\theta ^{2}}-%
\frac{\dot{F}}{3\theta }\right) \frac{k^{2}}{a^{2}}\eta  \nonumber \\
&&-\left( \frac{2\dot{F}\dot{\theta}}{\theta ^{2}}+\frac{2\ddot{F}}{\theta }+%
\frac{\dot{F}}{3}\right) \frac{k}{a}\mathcal{Z},
\end{eqnarray}%
where $\mathcal{X}^{f},\mathcal{X}^{pf},\mathcal{Z},\eta $ are respectively
the harmonic expansion coefficients of $\hat{\nabla}_{a}\rho ^{f},\hat{\nabla%
}_{a}p^{f},\hat{\nabla}_{a}\theta $ and $\hat{\nabla}_{a}\hat{R}$ (see for
example \cite{Lewis2000}). Moreover, here we are working in the CDM frame
(with the 'observer' comoving with dark-matter particles and so
free-falling) in which case we can set $A_{a}=0$ \cite{Challinor1999,
Lewis2000} to simplify computations. In this case we have (up to first
order)
\begin{eqnarray}
\hat{\nabla}_{a}\dot{F} &=& \sum_{k}\frac{k}{a}\dot{\epsilon}Q^{k}_{a},
\nonumber \\
\hat{\nabla}_{a}\ddot{F} &=& \sum_{k}\frac{k}{a}\ddot{\epsilon}Q^{k}_{a}.
\nonumber
\end{eqnarray}

The presence of the term $-\frac{27}{16}\frac{1-\frac{4}{3}\dot{F}\theta }{%
\theta ^{4}F_{G}}$ in Eq.~(\ref{eq:44}) is notable. As we have seen in the
above section, the magnitude of the dimensionless quantity $|\theta
^{6}F_{G}|$ is tiny at early times (deep into the matter- and
radiation-dominated eras) so that at that time this term dominates over the
other two in the squared brackets. If $F_{G}>0,$ then $-\frac{27}{16}\frac{1-%
\frac{4}{3}\dot{F}\theta }{\theta ^{4}F_{G}}\rightarrow -\infty $ at these
early times, which makes the perturbation $\epsilon $ unstable and it grows
quickly to spoil the linear theory. This is similar to the analysis of
stability in $f(R)$ models \cite{Song2007, Li2007}. For the subclass of
models with $F_{G}<0$, to which we are restricting ourselves, $-\frac{27}{16}%
\frac{1-\frac{4}{3}\dot{F}\theta }{\theta ^{4}F_{G}}\rightarrow \infty $ and
the value of $\epsilon $ quickly settles towards $-\frac{16}{27}\theta
^{4}F_{G}S$ . It can be checked easily that this is equivalent to $\hat{%
\nabla}_{a}F\doteq F_{G}\hat{\nabla}_{a}G$, which is as expected (note that
at these times the influence of $f(G)$ corrections is negligible and all
perturbation quantities except $\epsilon $ follow their standard GR
evolution). Note that this could be used as the initial condition for $%
\epsilon $ when we evolve it numerically, and the initial condition for $%
\dot{\epsilon}$ could be obtained simply by taking its time derivative.

This is not the whole story. We could use Eq.~(\ref{eq:19}) to substitute
the term $-3\frac{\dot{\theta}}{\theta ^{3}}\frac{k}{a}\mathcal{Z}-\frac{9}{2%
}\frac{\dot{\theta}}{\theta ^{4}}\frac{k^{2}}{a^{2}}\eta $ in Eq.~(\ref%
{eq:46}) to re-express the evolution equation of $\epsilon $ as
\begin{widetext}
\begin{eqnarray}
\label{eq:47} \ddot{\epsilon} + \left(\theta +
\frac{4\dot{\theta}}{\theta} \right)\dot{\epsilon} + \left[\left(1
+ \frac{4\dot{\theta}} {\theta^{2}}\right)\frac{k^{2}}{a^{2}} -
2\left( \dot{\theta} + \frac{\dot{\theta}^{2}}{\theta^{2}} +
\frac{2}{9}\theta^{2}\right) - \frac{27}{16}
\frac{1-\frac{4}{3}\dot{F}\theta}{\theta^{4}F_{G}}\right]\epsilon
&=& S'
\end{eqnarray}
\end{widetext}where
\begin{eqnarray}  \label{eq:48}
S^{\prime } &=&-\frac{3\kappa }{4\theta ^{2}}(\mathcal{X}^{f}-3\mathcal{X}%
^{pf})-3\left( \frac{\ddot{F}}{\theta ^{2}} -\frac{\dot{F}}{3\theta }\right)
\frac{k^{2}}{a^{2}}\eta  \nonumber \\
&&-\frac{9}{2}\frac{\dot{\theta}}{\theta ^{4}}\kappa \mathcal{X}^{f}-\left(
\frac{4\dot{F}\dot{\theta}}{\theta ^{2}}+\frac{2\ddot{F}}{\theta }+\frac{%
\dot{F}}{3}\right) \frac{k}{a}\mathcal{Z}.
\end{eqnarray}%
The term $\left( 1+\frac{4\dot{\theta}}{\theta ^{2}}\right) \frac{k^{2}}{%
a^{2}}\epsilon $ in Eq.~(\ref{eq:47}) makes the situation more complicated.
During the whole matter-dominated era and part of the
deceleration-to-acceleration transition period we have$\left( 1+\frac{4\dot{%
\theta}}{\theta ^{2}}\right) <0$. Thus, for the evolution of $\epsilon $ to
be stable on small scales ($k^{2}/a^{2}H^{2}\gg 1$), we must also require
that this term is subdominant compared with the third term in the squared
brackets; that is, $|F_{G}H^{6}|$ must be close enough to zero \emph{not only%
} at early times, \emph{but also} at low redshifts (\emph{e.g.}, $z\lesssim
\mathcal{O}(10)$). For example, in the $\Lambda \mathrm{CDM}$ limit $%
|F_{G}H^{6}|=0,$ so that the small-scale instabilities will never appear. In
general, since the deviation from $\Lambda \mathrm{CDM}$ in the $f(G)$ model
is roughly characterized by the deviation of $A$ from zero, $A\ll 1$ should
be satisfied in order that the model evades cosmological constraints from
linear spectra.

\begin{figure}[tbp]
\centering
\includegraphics[scale=0.9] {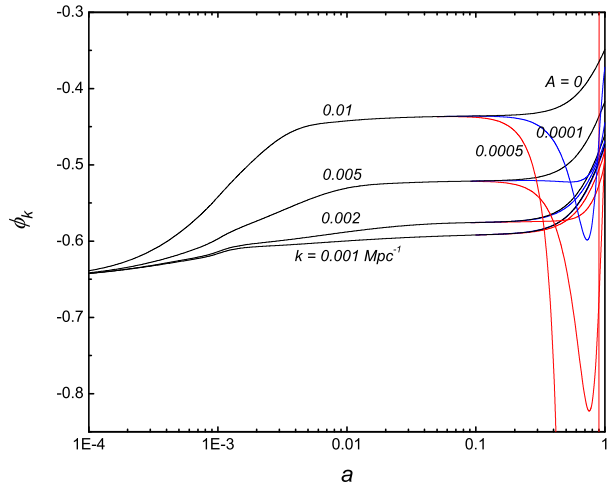}
\caption{(Color Online) Evolution of Weyl potential versus the
cosmic scale factor $a$ at different scales, $k = 0.001, 0.002,
0.005, 0.01\ \mathrm{Mpc^{-1}}$ respectively from bottom to top .
The values of $A$ are indicated beside the curves. $A = 0$
corresponds to the $\Lambda\mathrm{CDM}$ model.}
\label{fig:Figure3}
\end{figure}

\begin{figure}[tbp]
\centering
\includegraphics[scale=0.9] {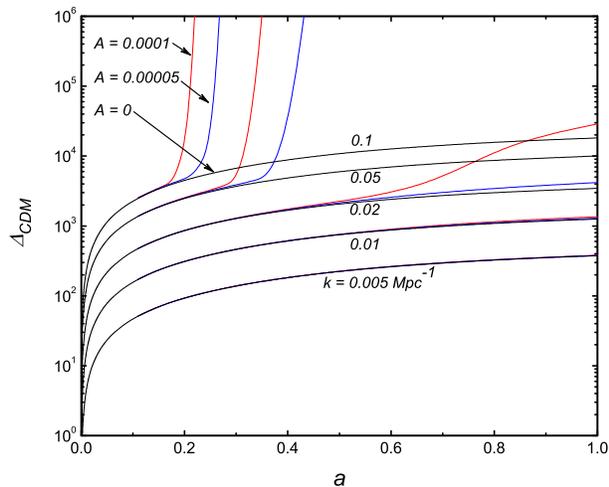}
\caption{(Color Online) Evolution versus cosmic scale factor $a$
of the cold dark matter density contrast $\Delta_{\mathrm{CDM}}$
at different scales, from bottom to top $k = 0.005, 0.01, 0.02,
0.05, 0.1\ \mathrm{Mpc^{-1}}$ respectively. The values of $A$ are
indicated beside the curves. $A = 0$ corresponds to the
$\Lambda\mathrm{CDM}$ model. Note that the rapid growths of
small-scale perturbations may make these scales leave the linear
regime much earlier than in the standard $\Lambda\mathrm{CDM}$
model.} \label{fig:Figure4}
\end{figure}

To illustrate the effects discussed above, we have shown the evolution of
some linear perturbation variables of the present model in Figs.~\ref%
{fig:Figure3} and \ref{fig:Figure4}. Plotted in Fig.~\ref{fig:Figure3} are
the evolutions of the (large-scale) Weyl potential $\phi _{k}$. This is the
coefficient of the harmonic expansion of $\mathcal{E}_{ab}$ as $\mathcal{E}%
_{ab}=-\sum_{k}k^{2}\phi _{k}Q_{ab}^{k}/a^{2}$ and is related to the
Newtonian potential, $\Psi ,$ by $\Psi =\phi -\kappa \Pi a^{2}/2k^{2}$ for
any specified $k$-mode, where $\Pi $ is the anisotropic stress. From Eqs.~(%
\ref{eq:11}, \ref{eq:33} - \ref{eq:36}), it is obvious that $\phi _{k}$
depends on $\epsilon $, and so from the analysis above it is easy to
understand why on smaller scales $\phi _{k}$ changes so dramatically. This
situation is quite similar to that in the Palatini $f(R)$ gravity model \cite%
{Li2006a} where $f(R)=R+\alpha (-R)^{\beta }$ with $\beta >0$. Since the
time evolution of $\phi _{k}$ determines the CMB power through the
integrated Sachs-Wolfe (ISW) effect as
\[
I_{l}^{\mathrm{ISW}}=2\int^{\tau _{0}}\phi _{k}^{\prime }j_{l}[k(\tau
_{0}-\tau )]d\tau ,
\]%
where $j_{l}(k\tau )$ are the spherical Bessel functions, and $\tau _{0}$
the conformal time at present, the extremely rapid variations in $\phi _{k}$
might greatly enhance the angular power spectrum of temperature
anisotropies, as in the the $\beta >0$ case of Ref.~\cite{Li2006a}. This
will provide the first stringent constraint on the present model (or equally
on $A$).

In Fig.~\ref{fig:Figure4} we have displayed the time evolution of the cold
dark matter density contrast, $\Delta _{\mathrm{CDM}},$ on different scales.
As expected, on large scales the $k^{2}$ term in Eq.~(\ref{eq:47}) never
becomes important and $F_{G}H^{6}$ is small enough for our choices of $A$,
so the deviation from $\Lambda \mathrm{CDM}$ ($A=0$) is small. On small
scales, however, the $k^{2}$ term is significant, and makes $\Delta _{%
\mathrm{CDM}}$ blow up quickly. This is also similar to the $\beta >0$
branch in Ref.~\cite{Li2006a}, and produces a matter power spectrum which is
strongly scale dependent for large $k$ (note the difference to metric $f(R)$
models \cite{Li2007}). This scale dependence will be strongly disfavored by
current data on galaxy power spectra (such as those from Sloan Digital Sky
Survey) and gives a second stringent constraint on the $f(G)$ cosmological
models. Considering these, although evaluating the numerical constraint on $%
A $ is beyond the scope of the present work, we can claim that the parameter
space for a viable $f(G)$ cosmology is highly limited. This, among others,
once more reveals the difficulties appearing in explaining the cosmic
acceleration with modified gravity models.

One should notice that the instabilities of the matter component found in
these $f(G)$, and in $f(R)$ models within the Palatini approach, are in fact
dependent on the nature of the dark matter and not only on the gravity
sector \cite{Koivisto2006}. For instance, instabilities are also found in GR
models where cold dark matter is coupled to a light scalar field \cite%
{amendola}. However, when dark matter is not cold, and so has a
free-streaming length, instabilities might go away, as occurs in some
interacting hot-dark-matter-dark-energy  models \cite{brookfield}. Hence,
ruling out these gravity models just due to the instabilities in CDM might
not be the last word.

\section{Conclusion}

\label{sect:Conclusion}

To summarize: in this work we have consider the cosmology arising from a new
modified gravity model, the modified Gauss-Bonnet model, where a function, $%
f(G),$ of the Gauss-Bonnet invariant is added to the Einstein-Hilbert
gravitational action to account for the current cosmic acceleration, at both
the background and first-order perturbation levels.

For the background-level evolution, we find that $f(G)$ models cannot
describe arbitrarily parameterized cosmic histories leading to the current
observations, because these histories are generally characterized by a
transition from $\dot{G}>0$ to $\dot{G}<0$ which might impose too strong a
requirement on the form of $f$.

For the linear-level evolution, we present the first-order perturbation
equations for the $f(G)$ gravity model in the CGI formalism with a
derivation which can also be applied to general $%
f(R,R^{ab}R_{ab},R^{abcd}R_{abcd})$ models. The special combination of $%
R,R^{ab}R_{ab}$ and $R^{abcd}R_{abcd}$ terms in the invariant $G$ ensures
that the set of perturbation equations resemble that in the $f(R)$ gravity
models and is much simpler than those in $f(R^{ab}R_{ab})$ or $%
f(R^{abcd}R_{abcd})$ theories. We analyze the perturbation equations and
find that only the $\partial ^{2}f/\partial G^{2}<0$ subclass of the models
could have stable perturbation growth. Furthermore, even within this
subclass, there will be a period during which the small-scale perturbation
growth is no longer stable unless $|F_{G}|H^{6}$ is close enough to zero.
This unstable growing period has two important consequences. Firstly, it
makes the gravitational potential $\phi _{k}$ change very rapidly, which may
greatly enhance the ISW effect and alter the CMB power. Secondly, the
small-scale dark matter density perturbations grow much more quickly than in
the $\Lambda \mathrm{CDM}$ paradigm, which might lead to a strongly
scale-dependent matter power spectrum. In both cases, the model therefore
faces stringent constraints from the cosmological data sets on linear
spectra. We thus conclude that the parameter space for a viable $f(G)$
cosmological model is highly constrained.

There are several points which arise from this work but are beyond the scope
of the present investigation. For example, it might be interesting to make
an analysis of the evolution dynamics for general $f(G)$ models and to
include the constraints arising from solar-system tests of gravity on the
model.

\begin{acknowledgments}
We thank Prof.~S.~Odintsov for communications stimulating this work and
Dr.~Yong-Seon Song for helpful discussion. The numerical calculation of this
work utilizes a modified version of the CAMB code \cite{Lewis2000, CAMB}.
B.~Li is supported by the Overseas Research Studentship, Cambridge Overseas
Trust and the Department of Applied Mathematics and Theoretical Physics at
the University of Cambridge. DFM acknowledges the Humboldt Foundation and
the Research Council of Norway through project number 159637/V30.
\end{acknowledgments}

\bigskip

\emph{Notes added}: There has recently appeared another work \cite%
{DeFelice2007} whose authors analyze a specific form of $f(R, G)$ theory by
different methods. They also conclude that their model does not give a
viable cosmology because of unstable behavior.



\begin{thebibliography}
\bibitem[Comment1(2007)]{Comment1} By taking the spatial derivative of the
relation $\nabla ^{b}R_{ab}=\frac{1}{2}\nabla _{a}R$, which results from the
contracted Bianchi identity, and using Eqs.~(\ref{eq:27}, \ref{eq:28}), one
can express $\Upsilon _{a}$ and $\Sigma _{ab}$ only in terms of $\hat{R}$
and kinematic quantities $\theta ,A_{a}$ (plus their derivatives). In this
way, the perturbation equations become higher ($4^{th}$) -order differential
equations for these quantities only -- as with traditional perturbation
calculations. However, we think that the CGI perturbation equations, when
are understood as propagation equations for the effective energy density,
heat flux and istropic and unisotropic stresses, make the physics more
transparent (and possibly easier to handle).
\bibitem[Comment2(2007)]{Comment2} Needless to say, failing to reproduce the
$\Lambda\mathrm{CDM}$ cosmic history does not necessarily mean the failure
of the model in general, as long as the background evolution can fit the
supernovae data well.
\end{thebibliography}
\end{document}